# Design, Implementation and Practical Energy-Efficiency Evaluation of a Blockchain Based Academic Credential Verification System for Low-Power Nodes

Gabriel Fernández-Blanco [1,2], Iván Froiz-Míguez [1,2], Paula Fraga-Lamas [1,2,*] and Tiago M. Fernández-Caramés[1,2]

1. Department of Computer Engineering, Faculty of Computer Science, Universidade da Coruña, 15071 A Coruña, Spain; angel.niebla@udc.es (Á.N.-M.); ivan.froiz@udc.es (I.F.-M.); tiago.fernandez@udc.es (T.M.F.-C.)
2. Centro de Investigación CITIC, Universidade da Coruña, 15071 A Coruña, Spain
* Correspondence: paula.fraga@udc.es; Tel.: +34-981167000

**Abstract:** The educational system manages extensive documentation and paperwork, which can lead to human errors and sometimes abuse or fraud, such as the falsification of diplomas, certificates or other credentials. In fact, in the last years, multiple cases of fraud have been detected, which have a significant cost to society, since they harm the trustworthiness of certificates and academic institutions. To tackle such an issue, this article proposes a solution aimed at recording and verifying academic records through a decentralized application that is supported by a smart contract deployed in the Ethereum blockchain and by a decentralized storage system based on Inter-Planetary File System (IPFS). The proposed solution is evaluated in terms of performance and energy-efficiency, comparing the results obtained with a traditional Proof-of-Work (PoW) consensus protocol and the new Proof-of-Authority (PoA) protocol. The results shown in this paper indicate that the latter is clearly greener and demands less CPU load. Moreover, this article compares the performance of a traditional computer and two SBCs (a Raspberry Pi 4 and an Orange Pi One), showing that is possible to make use of the latter low-power devices to implement blockchain nodes but at the cost of higher response latency. Furthermore, the impact of Ethereum gas limit is evaluated, demonstrating its significant influence on the blockchain network performance. Thus, this article provides guidelines, useful practical evaluations and key findings that will help the next generation of green blockchain developers and researchers.

**Keywords:** Blockchain; Distributed Ledger Technologies; Education; CV; Academic Forgery; Energy Efficiency; Decentralized Application; DApp

## 1. Introduction

The educational system manages multiple documentation and paperwork, which opens up the possibility of human errors and sometimes abuse or fraud, such as the falsification of diplomas, certificates or other credentials. Today is common to find news related to the academic fraud and to the rise of diploma mills [1]. It is also estimated than more than half of all people claiming a new PhD actually have a fake degree [2]. These incidents cause serious damage to society, especially because of the undeniable value of a university degree. Besides damaging the reputation of educational institutions, this type of fraud can cause problems of labor intrusion if the person presenting a false degree commits usurpation of functions that require a higher degree of training or specialization, as it occurs, for example, in the medical field.

However, the lack of efficient tools for background verification and the administrative bureaucracy, which are expensive and time-consuming, discourage the necessary consultation and verification tasks by the entities that are responsible for the veracity of the provided information.

Stakeholders like institutions (e.g., schools, universities), organizations, companies and individuals (e.g., students) face multiple language and administrative barriers in the verification of diplomas, degrees or other credentials, the translation and comparison of degrees at different educational institutions and/or sharing information between them due to the lack of interoperability. Moreover, there are no global standards accepted by educational institutions, which makes it difficult the evaluation of credentials in a homogeneous way and can cause issues both in academia and in the labor market. Therefore, it is necessary to have a trustworthy electronic verification of credentials to improve the verification and the validation process of knowledge/skills acquired during the educational process. In fact, it is important to develop a student-centered model that gives students ownership of their data and that allows them to control their academic identity and the information about them that an employer or other stakeholders can access. In addition to the complexity of current procedures and systems, the infrastructure where the student's information is stored is usually completely centralized in applications where no tracking or history of the student's grades is usually provided, so anyone with access to the system can make changes to the information (in many cases without leaving a trace).

Over the last decade, Distributed Ledger Technologies (DLTs), specifically blockchain [3,4], have been seen as an opportunity to address major problems affecting security, privacy and fraud control in education due to their features: decentralization, traceability, immutability, and authenticity. Therefore, the problem of counterfeiting diplomas, academic records (AR) or a curriculum vitae (CV) is a problem that can be addressed by using blockchain technologies [5–7].

A blockchain is a distributed record that is shared, replicated and synchronized by multiple computational nodes in a decentralized network [8]. Transactions that take place throughout the entire network are verified and entered into the registry by consensus of all participants, without the control of central authorities or intermediaries [9]. Such a consensus protocol, together with the use of asymmetric cryptography, results in a high level of information security and tamper resistance. Moreover, Decentralized Applications (DApps) are applications whose backend is executed in a blockchain, unlike traditional applications where the backend runs on centralized servers [10,11]. DApps acquire all the decentralization properties provided by the blockchain, resulting in applications that preserve data privacy and are transparent and resistant to Denial of Service (DoS) attacks [12].

Blockchains also introduced the concept of smart contracts, which are code stored in the blockchain that are executed after reaching certain conditions, without the need for trustful entities or third-party supervision [13]. In contrast to classical applications, the backend of a DApp relies on such contracts, which are immutable and indisputable. The security provided by a blockchain provides smart contracts with a great potential for implementing multiple practical applications [14], but they need to be secure against possible attacks [15].

As it will be later detailed in Section 2, there have been previous attempts to implement blockchain-based solutions for academic document verification. However, most of such solutions focus on the verification of academic certificates, not considering the continuous monitoring of the ARs. Such academic history can be tracked and verified, as it becomes clear that fraud can come through the student's career and not only at its end.

Although in recent years blockchain applications have become popular, they have also been criticized for having low performance, being too energy intensive or not decentralized enough [16]. The latest research has been devoted to the analysis of different techniques to benchmark [17] and optimize its operation [18], leading to what is called Green Blockchain. These novel techniques pave the way to mass adoption, as these would turn blockchain technologies as efficient as many centralized systems, besides benefiting from all the advantages provided by decentralization.

Considering the previously mentioned issues, this article describes a blockchain-based solution for verifying the authenticity of academic records and professional merits. Such a solution relies on a DApp that makes use of a decentralized database [19] based on the Inter-Planetary File System (IPFS) protocol [20], which stores the raw off-chain data. Such data identifies and tracks the academic information through the use of hashes that are stored in the blockchain [21].

The developed DApp can verify not only academic certificates, but also the complete academic career of a student. The low complexity and time required by the proposed solution to perform the verification avoids the vast amount of bureaucracy usually involved in this kind of validations. Moreover, not only these records can be verified after the issue of an academic certificate, but also before the student obtains his/her degree.

All the code of the developed DApp is available as open-source (under GPL-3.0 license), so it can be downloaded from GitHub [22] and then be used and/or modified by future researchers and developers.

It is important to note that a preliminary version of the work presented in this article was previously published as a conference paper [23]. That paper had a limited scope, focusing primarily on presenting a few experiments. In contrast, this article significantly extends that prior work, providing a thorough overview of the overall design, implementation and detailed validation. In summary, this article presents a blockchain-based solution with the following main contributions:

- It includes a thorough analysis of state-of-the-art systems that make use of blockchain for document verification.
- A detailed description is provided on the design and implementation of a novel blockchain-based system for AR and CV verification.
- A detailed evaluation of the proposed solution is presented in terms of performance (latency, throughput and gas limit) and energy consumption, when using in a traditional computer and two different low-power Single-Board Computers (SBCs).

The remainder of this article is structured as follows. Section 2 reviews the state of the art of traditional and blockchain-based systems aimed at document verification and academic anti-forgery, and describes the basics of blockchain. Section 3 presents the design of the proposed system, detailing the communications architecture, as well as the main functionality of the proposed system, whose implementation is described in Section 4. Next, Section 5 presents the results of the performed experiments, which were aimed at quantifying the system performance and its energy consumption when executing it in traditional and low-power computers. Finally, Section 6 summarizes the key findings of the experiments, while Section 7 is devoted to conclusions.

## 2. State of the art

*2.1. Traditional systems for document verification*

In the last years governments, companies and institutions have tried to achieve a high degree of trust in their certificates and fight counterfeit [24]. In fact, many universities around the world make use of their own systems to verify their diplomas and certificates. However, these systems are based on centralized models, which have well-known security issues:

- They introduce Single Points of Failure (SPoFs), so they are prone to DoS attacks, which impact data availability.
- They usually do not provide secure traceability, so, if an attacker manages to impersonate someone, he/she will be able to perform unauthorized modifications without leaving a trace.
- They imply the existence of central authorities, which are not necessarily trustworthy regarding the way they manage user private data.
- They are prone to data leaks, which is a critical concern in industries like healthcare, insurance or education.

Until recently, most stakeholders had to accept the previously mentioned risks as a trade-off for the easier and faster operation of centralized systems. However, consequences are becoming increasingly dangerous as companies gather more data [25].

*2.2. Blockchain basics*

DLTs like blockchains represent a novel approach to address the major drawbacks mentioned in the previous subsection. A blockchain is a sequence of timestamped blocks linked through cryptographic hashes [26]. In a blockchain, data are synchronized and shared among all nodes in a Peer-to-Peer (P2P) network [8]. The blockchain utilizes asymmetric cryptography, meaning each network node is assigned two keys: a private key for signing blockchain transactions, and a public key that acts as the user's unique identifier [27]. Once a node initiates a transaction, it signs it and broadcasts it to its peers. The transaction's signature serves for authentication and as guarantee of integrity (any data transmission errors prevent successful decryption). As the peers of the broadcasting node receive the signed transaction, they verify its validity before retransmitting it to other peers.

Transactions validated through this process and accepted by the network are organized and grouped into a timestamped block by specialized nodes usually known as miners. The selection of miners and the data included in the block are determined by a consensus algorithm (a more comprehensive definition is provided later in Section 2.3). Once a miner has compiled the blocks, they are broadcast back into the network. The nodes proceed to confirm the validity of the transactions within the broadcast block and ensure it references the previous block in the chain through the corresponding hash. If both requirements are met, the block is added to their chain, thus updating the transaction. Otherwise, the block is discarded. Thus, the described mechanism ensures the integrity of the stored data, as no one can modify the information without network consensus [9].

Blockchains can be categorized based on the type of data they manage, the availability of such data, and the actions a user can perform. This leads to a distinction between public/private as well as permissioned/permissionless blockchains [26]. It is worth noting that some authors use the terms public/permissionless and private/permissioned interchangeably. While this may be applicable in the context of cryptocurrencies, it does not hold true for other applications. In some cases, it is crucial to distinguish between authentication (i.e., who can access) and authorization (i.e., what can be done). However, these distinctions are still under debate, and their definition varies in the literature.

Public blockchains allow anyone to join without third-party approval, enabling them to act as a simple node or as a miner/validator. Economic incentives are typically provided to miners/validators in public blockchains such as Ethereum (one of the most popular blockchains) or Litecoin. In contrast, private blockchains limit network access. Many private blockchains are also permissioned, dictating which users can conduct transactions, execute smart contracts (self-executing code operated over the blockchain), or serve as network miners (for example, in a consortium blockchain where a predetermined group of nodes controls the consensus process). However, not all private blockchains are necessarily permissioned. For instance, an organization could establish a permissionless private blockchain using Ethereum.

Finally, it is also worth mentioning that blockchains can also be classified based on their intended use, with some solely designed for tracking digital assets, and others facilitating the execution of specific logic. Moreover, some systems employ tokens (e.g., Ripple, Ethereum), while others do not (e.g., Hyperledger, Corda).

A detailed explanation of blockchain inner workings is beyond the scope of this article, but the interested readers can explore further information in [26,28].

Considering the previously described features, Ethereum was selected for developing the anti-forgery system described in this article. Ethereum can be both deployed in a private or public blockchain and, being permisionless, it allows anyone to verify potentially forged documents. In addition, smart contracts enable automating the verification functionality through code written in a high-level language (Solidity). However, it must be noted that, for the proposed application, Ethereum has four limitations that had to be addressed by the solution proposed in this article:

- It is necessary to deploy an efficient and secure off-chain storage subsystem for storing a relatively large amount of structured data.
- It is necessary to study the restrictions related to parameters like gas to create an efficient solution.
- Since consensus protocols based on Proof-of-Work (PoW) have been traditionally energy hungry, it is necessary to analyze its power consumption and compare it with the latest Ethereum protocol (Proof-of-Authority (PoA)).
- To demonstrate the effectiveness of the overall solution, it would be ideal to try it when using low-power affordable devices as blockchain nodes.

*2.3. Consensus protocols for Ethereum and energy efficiency*

As it was indicated in the previous subsection, consensus protocols are used to reach an agreement among the blockchain nodes. The first consensus protocol made use of a Proof-of-Work (PoW) mechanism that was utilized by Bitcoin. Although PoW provides trust and security, its massive energy consumption brought in the last years discussions about its long-term sustainability [29–31]. For instance, The Cambridge Bitcoin Electricity Consumption Index (CBECI) [32] estimated that PoW consumption involved around 147 Terawatts-hour and 74 metric tons of $CO_2$ a year [33], which is the same consumption as a small country like Denmark. As a consequence, in the last years multiple new consensus protocols have emerged to maximize energy efficiency without losing the trustworthiness provided by PoW [34,35].

In general, the trade-off for increasing energy efficiency is to decrease decentralization. In practice, PoW cannot always provide complete decentralization either, since, when the network is not big enough, a few nodes with high mining capacity can monopolize the validation process.

In a permissioned scenario, the network is populated with nodes that have a certain degree of trust in each other. Therefore, there is no need to provide computational competition among the participating parties to ensure they are following the rules of the network. Hence, in permissioned blockchains, the use of an alternative protocol can accelerate block creation, since there is no need for keeping a high block time. Moreover, in this type of networks blocks can be larger than in a mainnet, resulting in a higher transaction throughput [36].

To tackle the previous issues different consensus protocols have been proposed in the last years to decrease energy consumption [37]. Proof-of-Stake (PoS) and Proof-of-Authority (PoA) are two protocols that provide a good balance between efficiency and performance without losing excessive decentralization. In addition, they are suitable for environments where the involved parties have similar interests (like the validation of ARs proposed in this article):

- Proof-of-Stake (PoS). It is the protocol currently supported by the mainnet of the Ethereum network and one of the most popular. There is a lot of research with respect to its energy consumption and, as demonstrated by the Crypto Carbon Ratings Institute [38], its energy consumption is orders of magnitude below Bitcoin's PoW mechanism. In PoS, validator nodes (i.e., the ones responsible for validating transactions and for creating new blocks) do not come to an agreement through the resolution of a computational problem as in PoW, but by making use of the amount of stake they possess. This eventually can lead to an imbalance on the validation process where the nodes with most stake are always selected, so parameters like randomness or coin age can be adjusted to keep the validation decentralized. Thus, PoS may involve new security risks, but it prevents other type of attacks, since an attacker needs to get stake to perform fraudulent transactions.
- Proof-of-Authority (PoA). PoA establishes a pool of known and trustworthy validator nodes. The guarantor element is not the amount of stake the validator nodes contribute with, but their public identity. This acts as an incentive to improve their standing on the network to keep them as validators. Moreover, validators need to comply with pre-requisites and regulations present in a smart contract. These rules and the lack of competition among the stakeholders provide a good level of trust. Furthermore, PoA is more resistant than PoW to 51% attacks [34], since it is way more complicated to control half of the nodes than half of the computational power. It is also worth noting that PoA provides fast transaction speeds and scalability, as it is possible to maintain a low number of validator nodes while the network grows. It is also energy efficient, as validator nodes are usually executed in dedicated hardware. However, in permissioned blockchains, PoA can be impacted by the acts of a malicious consortium among evil stakeholders. A practical example of the use of PoA is the Energy Web Chain (EWC) [39,40], a public open-source Ethereum-based platform that focuses in the energy industry that has demonstrated to be efficient and good-performing.

Finally, it is worth indicating that Ethereum's clients like Geth offer the possibility of alternating between a PoW algorithm (Ethash) and one based in PoA (Clique). Thanks to such a feature, in this article it was possible to compare PoA and PoW in a fair way through multiple tests, which are later detailed in Section 5.

*2.4. Blockchain for document verification*

Certificate forgery is a problem that has been previously studied by different blockchain-based solutions [41–43]. For instance, Tellew and Kuo proposed to manage healthcare training certificates through the use of smart contracts and a blockchain called CertificateChain [44]. Specifically, the authors explored the feasibility of storing full files into the blockchain by breaking them into slices. To demonstrate the feasibility of the proposed approach, CertificateChain was evaluated in terms of scalability and performance. The obtained results show that the operation of the system becomes unviable in mainnets (i.e., fully operational blockchain networks) where the size of the transactions is directly proportional to the transaction costs, but it can work well in private blockchains.

Regarding academic fraud, there are currently many studies and proofs of concept available, many of which were previously analyzed by Arno Pfefferling and Patrick Kehling in 2021 [45]. For instance, the Massachussets Institute of Technology (MIT) [6] developed an application where students' diplomas were stored, managed and verified through the Bitcoin blockchain. However, Bitcoin limits significantly the potential use cases of this kind of applications in comparison to other blockchains like Ethereum [13]. For example, Gresch *et al.* [21] took advantage of Ethereum and stored encrypted students' diplomas in its blockchain. In such a system the verification only consists in encrypting the provided diploma and then checking whether the resultant hash matches one of the hashes stored in the blockchain.

Turkannovic *et al.* [46] proposed a more global approach that aims to be used across European universities. The proposed system is based on the use of a token akin to the European credits (European Credit Transfer and Accumulation System, ECTS) and on a verification process that consists in checking the amount of tokens of a student. As the authors point out, the system could be further improved by introducing the standardization of the ARs among universities.

Chen *et al.* [4] proposed to use verifiable credits in cross-university courses. Thus, the credits earned by the students and the hash value of the records would be stored in a blockchain. Moreover, data would be maintained jointly by the universities, and one university would be capable of verifying the final examination of students, thus recognizing the credits from other universities, and at the same time, effectively protecting students' privacy using Elliptic Curve Digital Signature Algorithm (ECDSA). In the described system, the proposed architecture is conceived to make use of a consortium blockchain like Hyperledger Fabric, which can outperform other relevant solutions in terms of cost and security [47–49].

It is also worth mentioning that only a few previous works propose fully decentralized approaches. For instance, in the system described in [50], an administrator is responsible for encrypting the student's data, which are then uploaded to IPFS. These data need to be tamper-proof and not modifiable, so cipher texts are uploaded to a blockchain so their authenticity could be verified. The encrypted data can be decrypted and displayed to the student from IPFS using the proper credentials. Unfortunately, in [50] no implementation details are provided, and the system uses the mainnet of Ethereum, which significantly degrades performance and usability.

As a reference, Table 1 compares the features of some of the most relevant state-of-the-art blockchain-based solutions with the ones provided by the system presented in this article. As it can be observed in Table 1, there are other solutions aimed at verifying certificates and academic data, but only a few of them have been explicitly devised to verify the progression of the individual activities carried out by a student throughout a subject or a course. Moreover, although many systems support the storage of off-chain data, unfortunately such information is in general stored in centralized legacy systems (e.g., university local databases), which are prone to cyberattacks like denial-of-service (DoS) attacks, leaks or untraced modifications. Furthermore, most systems provide in their respective articles a good description of their inner workings, but only some of them quantify their performance and barely none considers energy consumption (specially for low-power devices). Finally, it is also relevant to point out that many of the platforms are not available as open-source software, although some of them have an official GitHub repository.

Therefore, after a thorough review of the state of the art, it was not found any blockchain-based solution that provided together all the features of the system presented in this article.

## 3. Design of the proposed system

*3.1. Communications architecture*

Figure 1 shows the communications architecture of the proposed system, whose main subsystems (highlighted in green in Figure 1) are:

- **Decentralized App**. The application enables users and professors to interact with the blockchain through a simple interface. Students can see the current state of their ARs, whereas administrators have the ability to update them. In addition, this subsystem provides an input field that can be used by any user that wants to verify an AR easily.
- **Blockchain**. It keeps the traceability of the students' ARs through their hashes. With this strategy, the amount of information stored in the blockchain is significantly reduced, but it requires keeping raw information outside the chain (off-chain). This increases transaction speeds and cost savings, since operational costs decrease (e.g., the gas spent on Ethereum-based applications is reduced).
- **Decentralized Database**. This database stores the private information of professors and students in a decentralized manner, as well as their public keys. The DApp interacts with this database in order to log in students and administrators, and allows for detecting new changes in the ARs. It is important to note that some decentralized databases like OrbitDB store the information in plain by default [52], so developers have to take privacy-protection actions to comply with laws regarding data privacy.

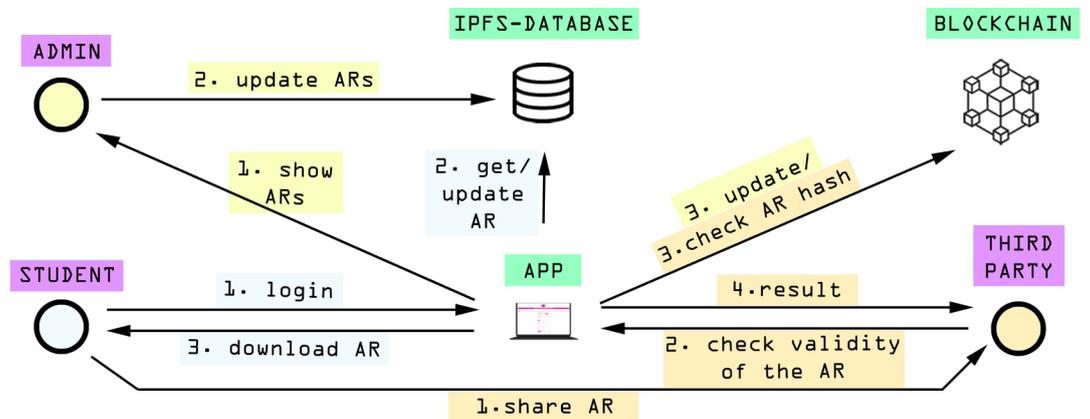

**Figure 1.** Relationships between actors and subsystems.

Figure 1 also shows the different actors that interact with the system (highlighted in purple), who are:

- **Students**. Once registered, students are able to see the evolution of their ARs after their periodic exams and courses. Whenever a student needs it, his/her AR can be downloaded (typically in a PDF file) and shared with a third-party, like an organization interested in hiring him/her that wants to validate the student academic merits.
- **Administrators**. After exams, professors register the marks obtained by the students in the decentralized database. However, such marks need to be verified before getting traced in the blockchain. Administrators can be seen as permissioned professors, who are responsible for looking into mark changes and update the hashes in the blockchain. Specifically, the developed DApp detects when the ARs state is changed by the professors and creates a list with the updated ARs, which will be shown to the administrator (e.g., to the head of the department or to a person that works for the university/school administration) and, if he/she approves the changes, he/she will create a transaction to update the AR hashes on the blockchain. Such an update can be performed individually for each AR or in batch (i.e., for several ARs at the same time) to accelerate the process. It is important to note that, as it was previously mentioned, each transaction only uploads the AR hash to the blockchain (not the AR raw data).

Table 1. Comparison of the proposed work with some of the most relevant state-of-the-art systems.

| Criterion / Platform | Verified Documentation | Verification Method | Verification of Progressing Activities | Blockchain | Consensus Protocol | Decentralized Off-Chain Storage | IPFS Support | Multi-Node Evaluation | Performance Evaluation | Low-Power Evaluation | Open Source |
|---|---|---|---|---|---|---|---|---|---|---|---|
| CertificateChain [44] | Healthcare training certificates | Permanent link to website | ✗ | Ethereum (designed for private networks, tested in Ropsten) | PoW | ✗ | ✗ | ✓ (up to 2 local nodes; tested in Ropsten) | Throughput | ✗ | ✗ (but all code is available in Zenodo and GitHub) |
| MIT Blockcerts [6] | Digital credentials | Through website | ✗ (no native support, but could be implemented) | Supports multiple blockchains | Depends on the used blockchain | ✗ (off-chain storage, but no native support for its decentralization) | ✗ (not native support, but can be implemented) | ✓ | ✗ | ✗ | ✓ |
| University of Zurich Blockchain (conceptual design) [21] | Higher education diplomas | Through website | ✗ | Ethereum | Not specified | ✗ (off-chain storage, but no native support for its decentralization) | ✗ | ✗ | ✗ | ✗ | ✗ |
| EduCTX [46] | Higher education credits | Requires to implement an API | ✓ (at credit level) | ARK | DPoS | ✗ (off-chain storage, but no native support for its decentralization) | ✗ | ✗ | ✗ | ✗ | ✓ |
| Chen et al. (conceptual design) [4] | Academic records (at activity level) | Requires to implement an API | ✓ | Hyperledger Fabric (proposed) | Not indicated | ✓ (proposed) | ✗ | ✗ | ✗ | ✗ | ✗ |
| Sultana et al. [50] | Academic certificates | Through website | ✗ | Ethereum | PoW | ✓ | ✓ | ✓ | Throughput, latency | ✓ | ✗ |
| PriFoB [51] | Digital credentials | Requires to implement an API | ✗ (no native support, but could be implemented) | Proprietary public permissioned | PoA | ✗ (off-chain storage, but no native support for its decentralization) | ✗ | ✓ | Throughput, latency | ✗ | ✗ (but code available in GitHub) |
| This Work | Academic records (at activity level (e.g., exame, lab)) | Through website | ✓ | Ethereum | PoW and PoA | ✓ | ✓ | ✓ (up to 16 local nodes) | Performance, latency, gas limit | ✓ | ✓ |

- **Third-party**. It is any external user that wants to verify the validity of an AR. This is simply performed by uploading the provided AR document to the DApp, which will show almost instantly whether the introduced AR is part of the blockchain or not (by comparing its hash with all of the hashes stored in the blockchain).

To deploy the devised architecture, every participating entity needs a full node to interact with the blockchain. However, the process of setting up a full node is not straightforward for most users. As this would not favor the adoption of the proposed solution, there are other ways to access the blockchain such as using node providers. For instance, there are third-party intermediate services that provide already configured full nodes able to interact with the DApp and with the blockchain network [53]. In the case of Meta- mask wallets [54], they already use these services to emit transactions, so simply using a Metamask wallet hides the complexity of the infrastructure to non-technical users.

Figure 2 provides a low-level perspective of the three main subsystems of the DApp and how they communicate. Such a Figure depicts an architecture that follows a Remote Procedure Call (RPC) model, where the DApp is the client and the RPC node is the server. The client just sends requests to the RPC node, allowing an easy real-time interaction with the blockchain data.

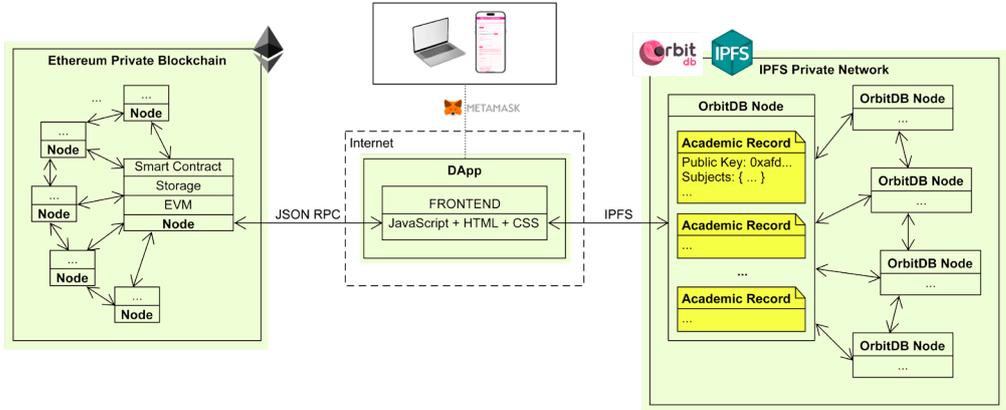

**Figure 2.** Main subsystems of the proposed DApp and their interconnections.

*3.2. Main functionality of the system*

3.2.1. Zero-Knowledge Proof authentication

In order to describe how user authentication is performed, it must be first indicated that the state of an Ethereum blockchain consists of accounts and contracts. User accounts are represented by private (SK) and public keys (PK), which are necessary to interact with the blockchain. Wallets allow users to easily generate accounts to sign messages, hold assets or perform transactions in the network.

To access an AR, it is necessary to log in into the DApp. Among the different techniques to log in users into Web3 applications, the developed application implements a Zero-Knowledge Proof (ZKP) based login system that involves a signature verification using an Elliptic Curve Digital Signature Algorithm (ECDSA) [55] to determine that the PK/SK holds the legitimacy of the user. For such a purpose, the user needs to sign a string with his/her SK, resulting in a ciphered string. Then, to make sure that the user is the legitimate holder of the SK, he/she needs to introduce the aforementioned string, the cipher string and his/her PK. With these three values, the DApp can determine whether the user used his/her SK to sign the message, without actually knowing or storing the SK.

It is worth noting that in the version of the developed system available in the public open-source repository [22], the user has to type the text to be signed, but this could be easily automated by providing a randomly generated message.

3.2.2. Access to the ARs

All students have their data stored in the database, so their ARs are linked to their PKs. However, initially, AR hashes are not included in the blockchain. For such a purpose, every student has to ask for an initial transaction to get his/her AR hash tracked by the blockchain the first time they login into the DApp. Once a DApp administrator performs such an initial transaction, the resulting transaction hash (not to be confused with the AR hash) will be appended to the student AR in the database. Thus, after the initial transaction the AR can be displayed to the student upon request.

For the sake of clarity, the previously described process is illustrated in the sequence diagram shown in Figure 3, where OrbitDB is the decentralized database.

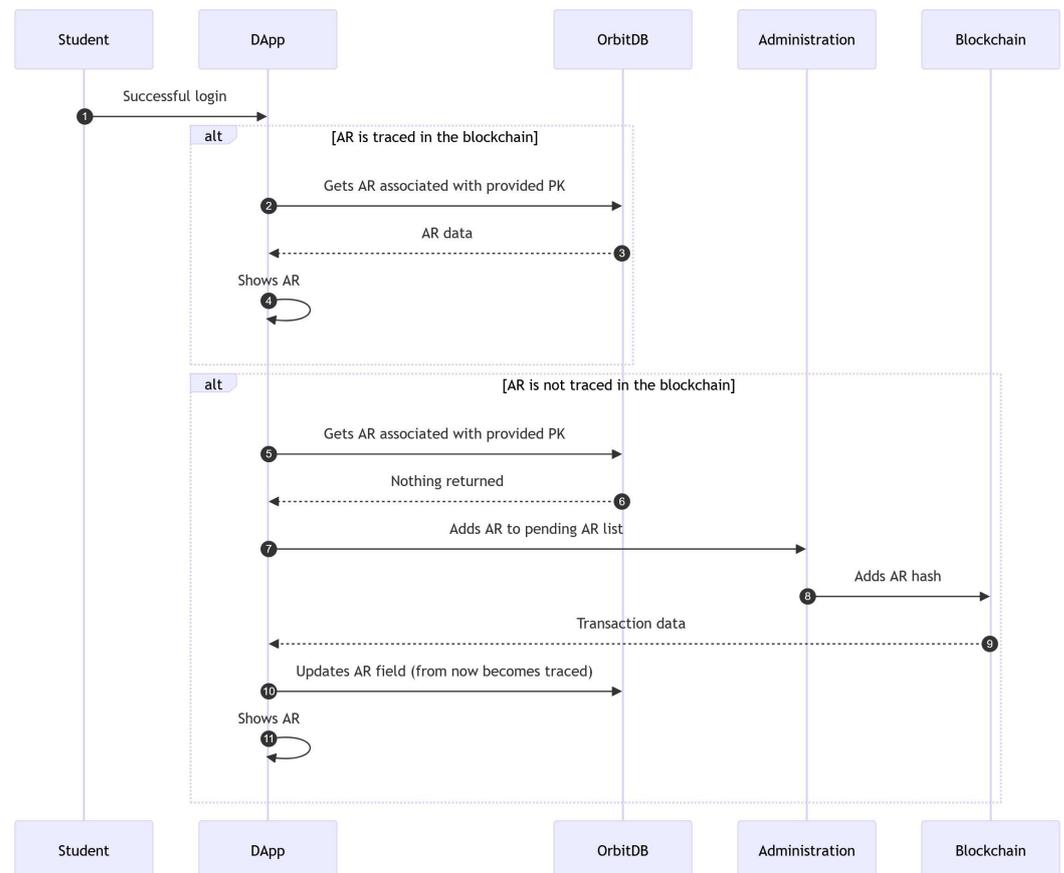

**Figure 3.** Issue of an AR for the first time.

3.2.3. AR update

After every exam period, professors update student ARs with the corresponding marks. These changes need to be confirmed by a group of administrators. Even though the implementation provided in the public open-source repository [22] just uses one administrator account for demonstration purposes, it is necessary more than one for the update process in a real deployment. The more administrators validating these updates, the less the chances of tampering or bribery.

The administrators will have access to all the modified ARs of the database. Once checked by the administrator, a transaction is performed to update the ARs stored in the blockchain (i.e., the ARs hashes are updated). All blockchain transactions are public, so anyone with access to a web browser-based block inspector can see the transaction information [56].

All the previously described updating process is detailed step by step in the sequence diagram shown in Figure 4.

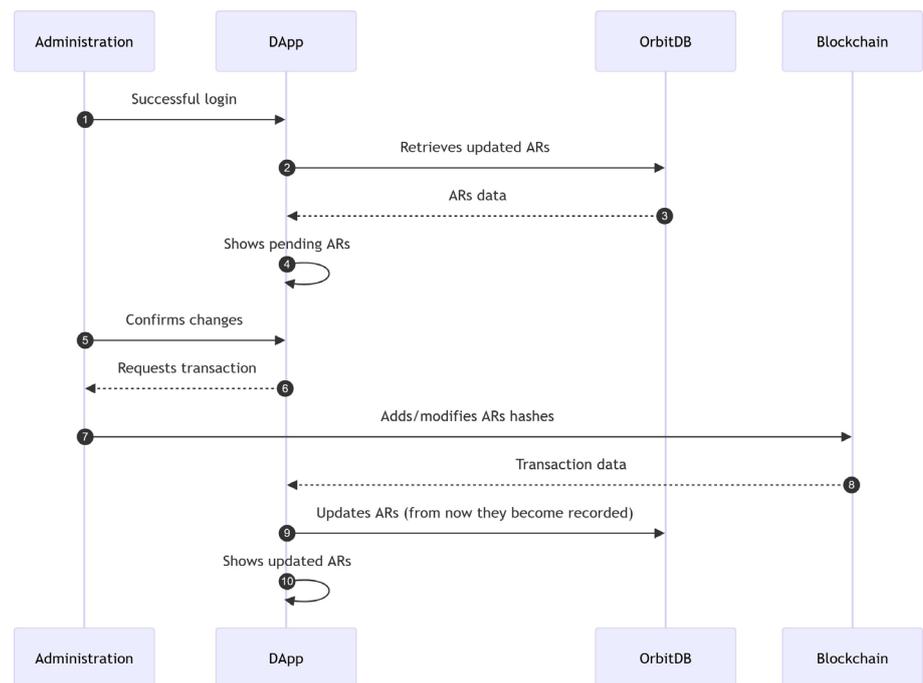

**Figure 4.** AR update process.

3.2.4. AR verification

Students can download and share their ARs with any third-party in order to prove their merits. As it is illustrated in Figure 5, the verification process simply takes the AR file (introduced by the third-party) and checks whether its corresponding hash is stored in the blockchain. If such a hash matches any of the stored hashes, then the system can confirm that the AR is tracked by the blockchain, therefore it is valid and tamper-proof.

## 4. Implementation of the system

### 4.1. DApp

The main functionality of the DApp described in Section 3.2 was implemented as follows:

- Authentication. The application first asks the user to verify his/her identity through his/her Metamask wallet (illustrated in Figure 6). Then, users can sign the messages or the transactions that will be submitted to the network (an example is shown in Figure 7).
- AR access/first issue. After a successful PK/SK verification, the application will look for the student's PK in the off-chain storage. Note that, as it was previously mentioned, the student's data must be already present in the database (i.e., the AR database is indexed by the PKs, as they are unique values). Once found, the application checks the transaction hash of the student's AR. If the transaction hash field is not empty, then the AR is already recorded in the blockchain, so the application can display the AR to the student directly (an example of a retrieved AR record is shown in Figure 8). On the contrary, if the transaction hash field is empty, that means that the AR is stored in the database, but it is not recorded in the blockchain. Therefore, the administration would perform a transaction to trace it in the blockchain for the first time.

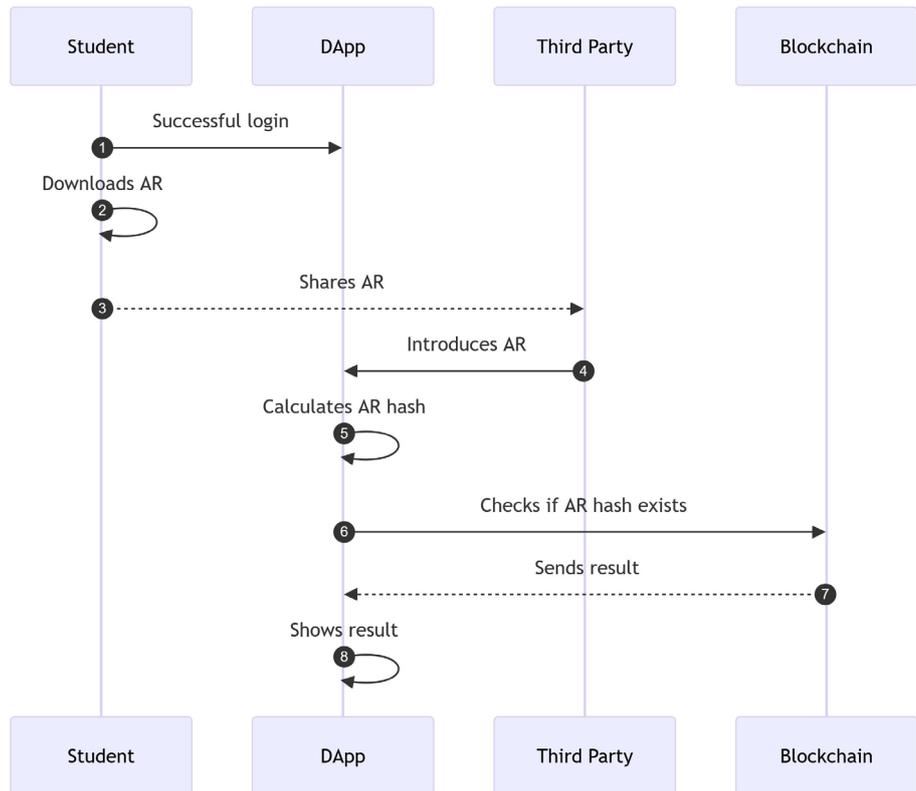

**Figure 5.** AR verification process.

- AR update. Any slight modification on the AR file will completely change its hash. Therefore, a new hash represents an AR update (or a new AR) that is not traced in the blockchain. When this happens, the application adds the AR (and all the updated ARs) to a 'pending AR list'. This list of ARs is what the administrator will see on his/her interface (an example is shown in Figure 9), waiting to be confirmed after the validation process.
- AR verification. As it was previously mentioned, a student can download his/her AR to share it with third parties, typically, an entity that wants to verify a CV. Such a downloading can be performed through a link that is enabled once the student has gained access to his/her information (such a link can be observed at the top of Figure 10). Then, the mentioned third-party would validate the CV merits by introducing the AR file into the application through the menu shown in Figure 11. The verification process is very fast (just a few seconds), as the application only needs to encrypt the AR (using the keccak256 function) and then to check whether its hash matches one of the hashes stored in the blockchain.

### 4.2. Smart contract and decentralized storage

The core of the blockchain-based system relies on a smart contract that makes use of two essential functions. The first one is the 'store' function, which introduces the AR hash

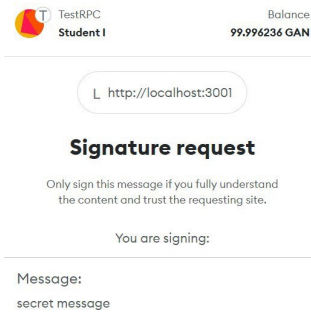

**Figure 6.** Metamask signature request in the DApp.

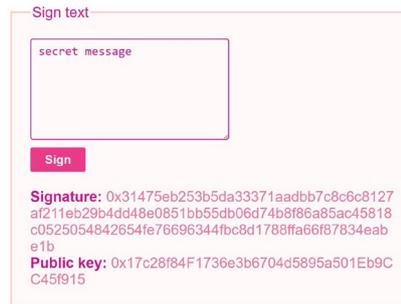

**Figure 7.** Message signed successfully with the developed DApp.

Id: 0x4ac13fc45076525bd5e883fbf8cd4ed40cf9b251ae9ee
ddf7e5a0e784f2438f6
Public Key: 0x17c28f84F1736e3b6704d5895a501Eb9CC45
f915
Degree: Computer science
Academic title issue date:
Name: Rose
Surname: Howard

| Subject | Mark | Subject Type | Course |
|---|---|---|---|
| Computing Theory | 5.2 | Basic Core | 23/24 |
| Calculus | 2.7 | Basic Core | 23/24 |
| Bussiness Management | 6 | Basic Core | 23/24 |
| Programming I | 5.7 | Basic Core | 23/24 |
| Discrete Math | 8 | Basic Core | 23/24 |
| Principles of Computer Engineering | | Basic Core | 23/24 |
| Programming II | | Basic Core | 23/24 |
| Linear Algebra | | Basic Core | 23/24 |
| Physics | | Basic Core | 23/24 |
| Statistics | | Basic Core | 23/24 |

**Figure 8.** Example of a retrieved AR shown by the DApp, where several subjects are not yet be evaluated.

to an array. The second one is the 'check' function, which verifies whether the submitted hash is present on the array or not.

OrbitDB [19] was selected as decentralized database. It is based on IPFS and provides an easy way to store and retrieve information through Create, Retrieve, Update, Delete (CRUD) operators. The decentralized database operates in a similar way to a blockchain, so every peer in the network has a synchronized view of the database. Moreover, the stored data are scattered among the participating peers and are tracked by an address and a hash table. OrbitDB offers developers five predefined types of databases (log, feed, keyvalue, docstore and counter), each with its own API and CRUD operators. The docstore database stores information with JSON format, so it is a good fit for managing structured data. OrbitDB databases also provide a control access system to manage the read/write permissions of the users.

It must be indicated that OrbitDB performance can be tested in terms of response time. Although such an evaluation is out of the scope of this article, the interested reader has further information in [57], where OrbitDB was used for a healthcare DApp. In such a paper, OrbitDB nodes were deployed locally and remotely (in Fog and Cloud computing environments), and it was found that response times were in the order of milliseconds, which demonstrates that OrbitDB is suitable even for scenarios where response time is critical.

**Figure 9.** Pending AR list shown to the administrator.

**Figure 10.** After the updating process, students can see the added marks.

**Figure 11.** Successful verification of an AR.

## 5. Experiments

### 5.1. Experimental setup

The devised experiments were carried out to evaluate the performance of the proposed system in terms of query latency, transaction throughput, resource usage and energy consumption. It is worth pointing out that the aim of the tests was not to provide an extensive analysis on every factor that impacts blockchain performance or efficiency [36,58,59], but to show the overall performance of the system in a realistic but limited permissioned scenario. In addition, the tests were also aimed at evaluating the performance of the system in resource-constrained devices, where power consumption is crucial [60].

Specifically, the following parameters are evaluated in the following subsections:

- Latency of read-operations, which allow for determining how fast the developed system responds to requests such as AR verifications.
- Throughput of write-operations, which enables analyzing the speed of the blockchain implementation.
- CPU usage and CPU power consumption.
- Energy consumption when deploying a node in two different resource-constrained devices.

For the sake of brevity, the performed experiments were focused on evaluating the power consumption of the consensus protocol, which usually supposes the most energy-hungry subsystem of a DApp.

The experiments were carried out in a local network, so it was necessary to use a client framework. Such a framework allows for configuring and for executing a node able to access the network. The used framework was Geth, which was selected due to its flexibility and good documentation. However, it is worth noting that there is previous research on other popular alternative frameworks that focus on performance and efficiency [61,62], which are interesting for real-world applications that make use of resource-constrained devices. For example, Nimbus [63] and Reth [64] are good alternatives for Green Blockchains, since they provide a consensus and an execution client focused on energy efficiency and Internet of Things (IoT) devices.

*5.2. Latency of read-operations*

In order to measure the latency of read operations, Flood [65] was used to 'flood' the developed Ethereum-based application with read-only RPC requests. The latency of each request was measured as the time that goes by between the submission of a transaction and the confirmation response that it is valid. It must be noted that latency is one of the metrics that potential users are usually more concerned about when deploying a DApp, essentially due to its impact on the overall performance and on user experience. The higher the number requests per second (rps) are emitted, the higher the load of the RPC node (i.e., the server) in terms of CPU cycles, I/O speed and memory usage.

To create a realistic scenario, sixteen nodes were used for this set of tests. The selected read-only operation for the tests was "eth_call". As a reference, two out of the sixteen nodes were evaluated (Node1 and Node2). Tests were performed for 10, 100, 500, 1000, 1500 and 2000 rps with a duration of 60s (i.e., for each rate, the tool maintained a specific pace of requests per second for 60 seconds). The results for each rate were obtained in ascending order according to the number of rps and then three percentiles were calculated (p50, p90 and p99).

The obtained results are shown in Figure 12. As it can be observed, Figure 12 has a sort of U-shape for each curve, having more latency at 10 rps than for 500 rps. This behavior should be ignored, as it does not represent realistic nor reliable outputs for this range. Since such a reported issue is inherent to the used test tool (Flood), it has been preserved in the Figure so that future researchers can consider it. However, from 500 rps the nodes start behaving as expected, having a smooth increase of the latency as rps increase.

Moreover, it can be observed in Figure 12 that, as expected, the lowest latency can be found for 500 rps, which is a realistic rate in practical scenarios. Furthermore, for the worst case evaluated (i.e., for a maximum of 2,000 rps), latency is still really low (in the order of milliseconds). Loads above 2,000 rps will derive into a relevant increase in latency, but such loads are not usual in a typical academic data verification application.

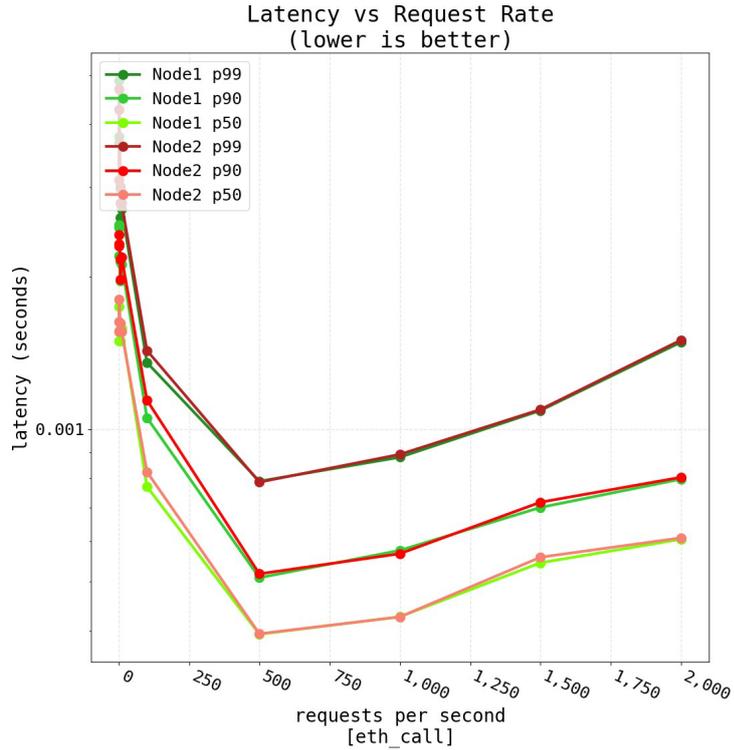

**Figure 12.** Latency results when performing up to 2,000 transactions of a typical read operation.

*5.3. Throughput of write-operations*

For evaluating write-operation performance, Pandoras [66] was used as stress testing tool. Such a tool is able to automate sending transactions between externally-owned accounts. The experimental results were recorded through Geth, measuring different metrics to analyze node performance and memory/CPU consumption.

The throughput ($T$) of write-operations was calculated as the number of successful transactions per second (TPS), defined by:

$$T = n_t/t_t \tag{1}$$

where $n_t$ is the total number of successful transactions and $t_t$ is the total time in seconds. Such a total time refers to the time required to create the blocks that contain $n_t$ transactions that will be appended to the blockchain (i.e., the workload handled per unit of time).

The network was set up with 16 nodes and 4 validators, configuring the gas limit to 1 million units of gas. Regarding the options provided by Pandoras, the transactions were set to EOA type, and the batch size was set to 1 (i.e., one transaction at a time). A fixed number of transactions was emitted (250, 500, 750 and 1,000) and the resulting transactions per second (TPS) were averaged 100 times per point.

The first set of tests measured the TPS of the network using PoW and PoA, when both made use of the same network parameters. The obtained results are shown in Figure 13. As it can be observed, PoA performs better with a moderate number of emitted transactions (up to 1,000). As the number of transactions increases, the differences are reduced significantly, mainly because of setting the gas limit to 1 M units, which becomes rapidly a bottleneck for the throughput of the network.

The influence of the gas limit was analyzed in a second set of tests that explored how TPS evolved when using different gas limits (from 1 to 60 million units of gas), while keeping the same consensus protocol (PoA) and network parameters unchanged. The obtained results are shown in Figure 14, where it can be observed that increasing the gas limit derives into increased TPS. However, there is a point in each curve where transaction throughput stops increasing, which is mainly due to the number of emitted transactions and to the impact of other blockchain parameters. In any case, in a realistic scenario, a gas limit of 1 M units of gas should be enough to deal with an eventual complex data manipulation for this application (as of writing, the public network of Ethereum uses 30 M as gas limit).

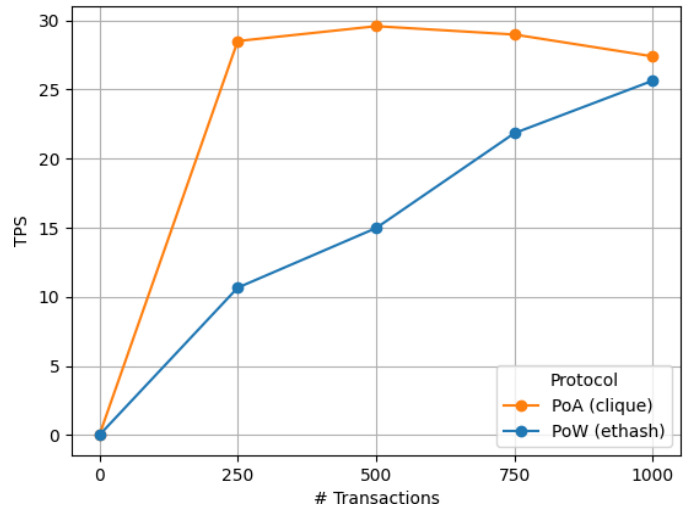

**Figure 13.** Transactions per second of the network with PoW and PoA.

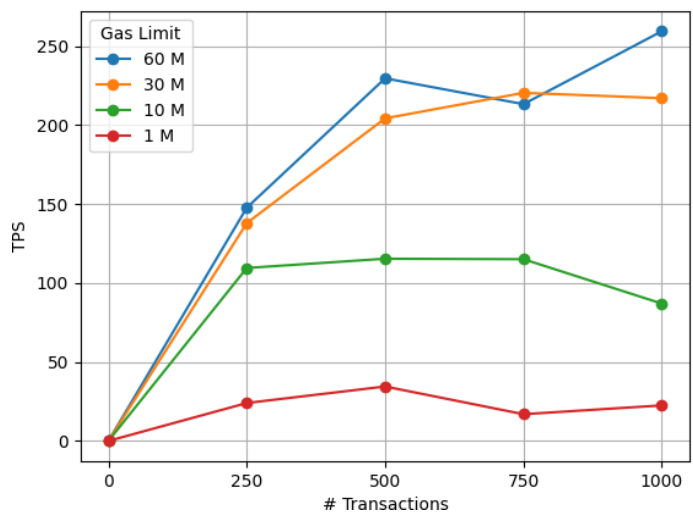

**Figure 14.** Transactions per second for four different gas limits using PoA.

*5.4. CPU usage and CPU power consumption*

A set of tests was carried out to compare the CPU usage and the energy consumption for PoW and PoA. For such a purpose, a blockchain network was deployed with 24 nodes, 3 of them acting as validators/sealers. Such an amount of nodes was selected to recreate a realistic scenario for the developed DApp.

These tests considered two scenarios. The first one was aimed at evaluating the network resource consumption at idle state (i.e., with no transaction load, just block mining/sealing). On the other hand, the second scenario put under stress the network by launching different amounts of transactions (between 100 to 1,000). CPU usage was measured using the Geth built-in metrics. Note that Geth measures CPU Usage as the percentage of use of one processing core (i.e., 100% means that one processing core is fully dedicated). The electrical consumption was measured using HWiNFO [67], a well-known software that provides real-time CPU information. More precisely, the metric recorded by this tool was "CPU Package Power", which provides the total energy consumed by a CPU.

Figure 15 shows the CPU usage for the first scenario (i.e., under no load). As it can be observed, even in idle PoW clearly requires much more CPU usage than PoA. However, such a difference between PoW and PoA is narrower in terms of power consumption, as it is shown in Figure 16, although, overall, PoA has a lower CPU power consumption.

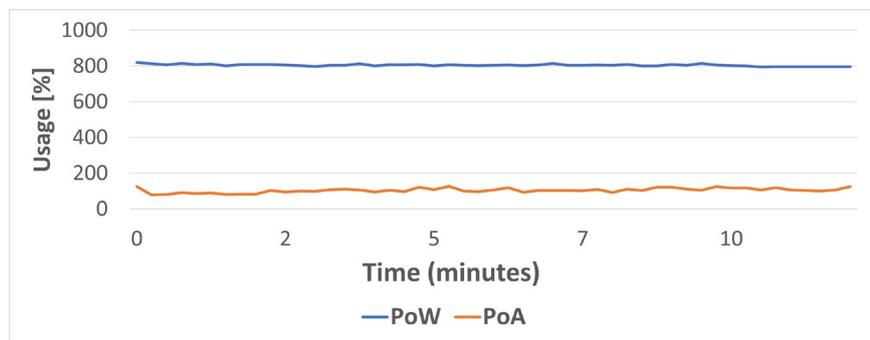

**Figure 15.** CPU usage in idle for PoW and PoA.

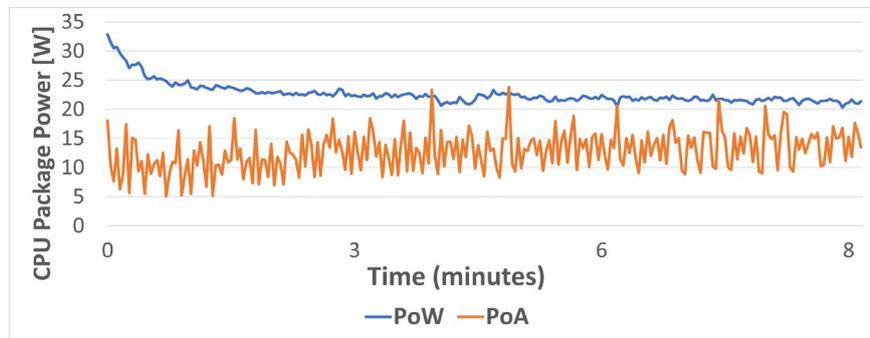

**Figure 16.** CPU power consumption in idle for PoW and PoA.

When the network is put under stress, the obtained results differ, as it is shown in Figures 17 and 18. As it can be observed, PoA CPU usage is clearly lower than for PoW (specifically, PoW average CPU usage is 66% higher than for PoA). However, during very specific high-transaction load peaks, power consumption is higher for PoA than for PoW. In fact, PoW average power consumption is only 23% higher than for PoA due to such peaks. In any case, it is important to take such a power consumption percentage with caution, since CPU Power measurements are not performed in an isolated way, so other processes can be executed in parallel by the CPU (e.g., by the operating system) and therefore can impact the obtained results.

*5.5. Practical evaluation on resource-constrained devices*

One of the foundations of DLTs are distributed communications. In this aspect, one of the most popular current trends consists in performing data and request processing close to the end user (e.g., Edge or Fog Computing [68]), thus offloading part of the workload of the cloud servers. Moreover, due to the need for deploying a large number of devices in certain applications, Single-Board Computers (SBCs) are commonly used due to their small size, reduced power consumption and low cost.

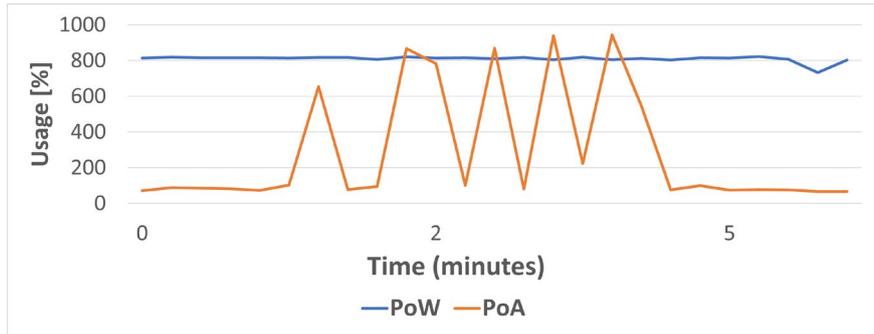

**Figure 17.** CPU usage for PoW and PoA under stress.

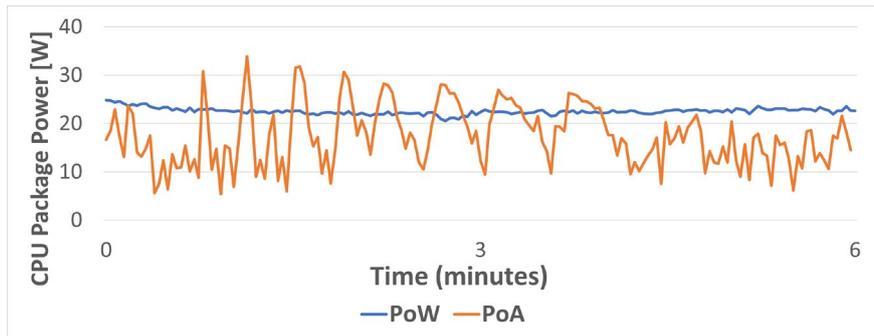

**Figure 18.** CPU power consumption for PoW and PoA under stress.

5.5.1. Response latency

The performed latency tests allowed for determining the maximum request load that each SBC can handle without degrading response time. Figures 19 and 20 show the read-operation latency obtained for two different SBC devices (Orange Pi One and Raspberry Pi 4, whose main specifications are shown in Table 2) under different workloads. Specifically, Figure 19 shows the results obtained for the Raspberry Pi 4 for up to 1,000 requests per second (with a duration of 60 seconds per each test), while Figure 20 shows the read-operation latency obtained for the Orange Pi One when performing up to 50 requests per second.

| Board | SoC | CPU Clock (GHz) | Memory (GB) | Ethernet |
|---|---|---|---|---|
| Raspberry Pi 4 | Quad-Cortex A72 (ARM64) | 1.8 | 1 + 1.1 Swap | Gigabit |
| Orange Pi One+ | H6 Quad-core A53 (ARM64) | 1.8 | 1 + 1.1 Swap | Gigabit |

**Table 2.** Main characteristics of the tested SBC hardware.

As it can be observed in Figure 19, the Raspberry Pi 4 offers response times similar to those shown in Figure 12 up to 500 requests per second, despite being the latter Figure obtained on a higher performance computer. In the case of the Raspberry 4, after the initial spike (preserved in the Figure for the sake of transparency, but due to the behavior of the used measurement tool, as it was previously described in Section 5.2), latency remains under 10 ms up to 600 requests per second. After such a number of requests per second, latency increases significantly with the number of requests except for the 50th percentile, where it remains low throughout all tests.

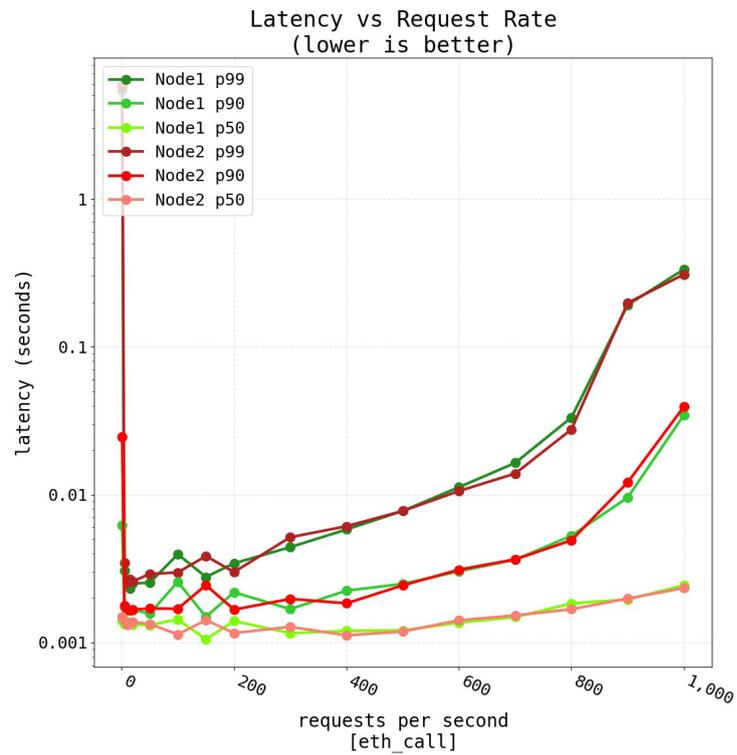

**Figure 19.** Latency results when performing up to 1,000 transactions on the Raspberry Pi 4.

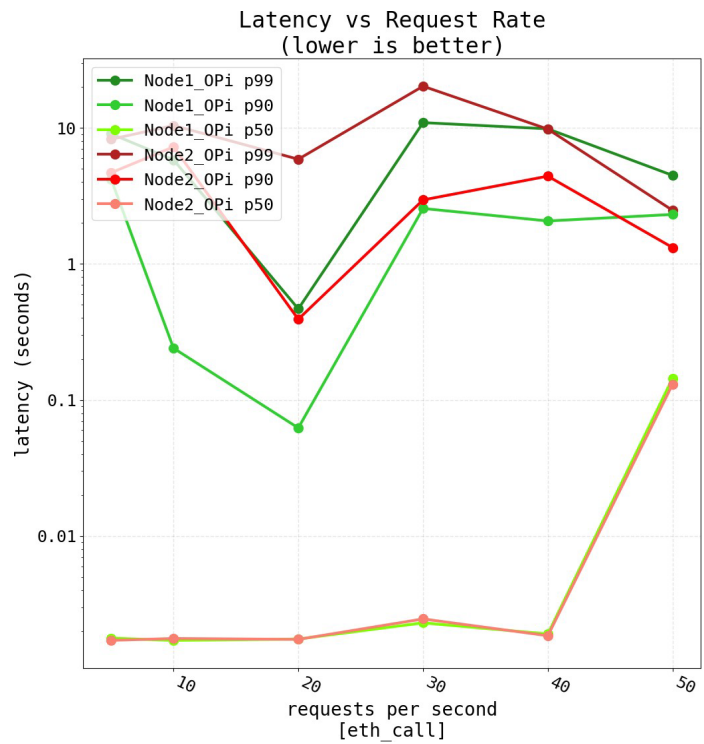

**Figure 20.** Latency results when performing up to 50 transactions on the Orange Pi One.

In the case of the Orange Pi One, due to the hardware differences with the Raspberry Pi 4, response times are clearly higher, as it can be seen in Figure 20. For such a SBC, the maximum number of evaluated requests per second was 50 since the latency was notably high, reaching values of up to 10 seconds. In the case of the 50th percentile, it remains stable under 10 ms until 40 requests per second, exceeding 100 ms for 50 requests per second.

5.5.2. Network performance

The Raspberry Pi 4 was selected for carrying out network performance tests as previously described in Section 5.3. This was due to the fact that the Orange Pi, although it was able to run both consensus protocols, it did not have enough power to perform high-load stress tests.

Thus, Figures 21 and 22 show the TPS obtained under different workload levels for the Raspberry Pi 4 in a network of four nodes and one validator. Specifically, Figure 21 shows the performance of the Raspberry Pi 4 for PoA at different gas limits. For obtaining such a Figure, the same number of transactions employed in Section 5.3 was sent but each point was averaged only 20 times due to the notable difference in hardware performance. In addition, it should be indicated that, when using resource-constrained devices, the gas limit cannot be too high: a high gas limit allows for a greater allocation of computational resources, thus avoiding network congestion, but due to the lower computational capacity of the SBCs, the synchronization process becomes slower when more transactions are validated for each block. In contrast, such a computational limitation implies that setting the gas limit too low may cause transactions to run out of gas due to their intrinsic complexity.

For such reasons, during the tests, the minimum gas limit was set to 5 M and the maximum to 30 M. The obtained results show a similar behavior to the ones depicted in Figure 14, but with a lower TPS performance, demonstrating that increasing the gas limit leads to an increase in TPS.

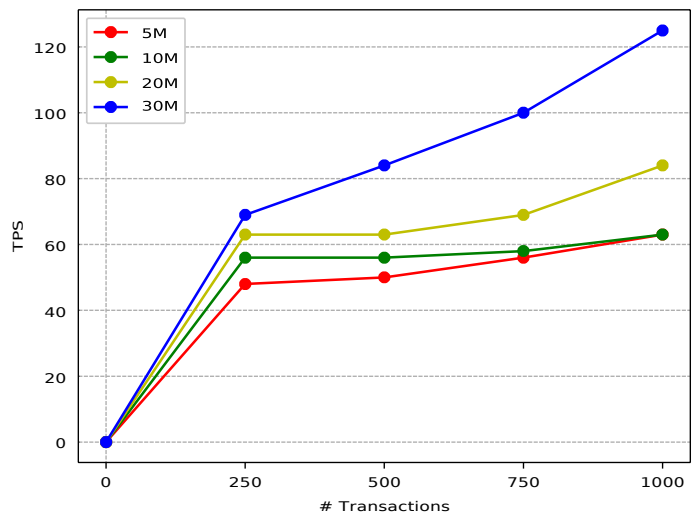

**Figure 21.** Network performance for PoA with different gas limits on the Raspberry Pi 4.

Figure 22 shows the difference in performance of PoW and PoA when the previously described test network using a gas limit of 20 M. Like in the non-SBC based case, the Figure shows, for different amounts of transactions, the remarkable difference in TPS between PoA and PoW, thus indicating that, under the selected experimental conditions, PoA is a better fit for the evaluated SBCs.

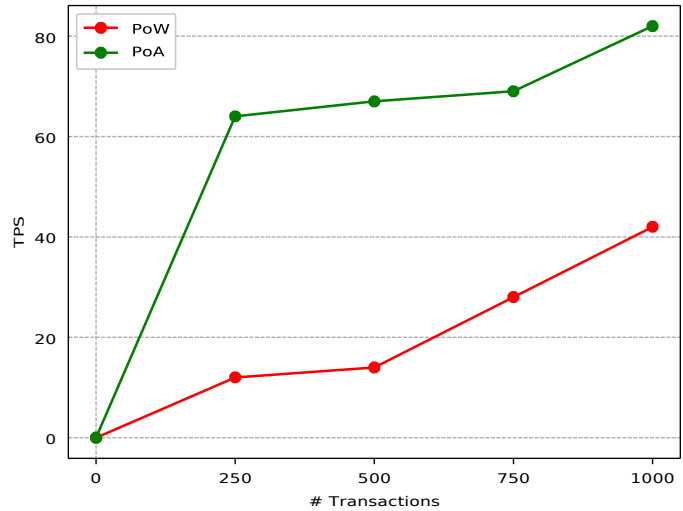

**Figure 22.** Network performance for PoW and PoA for a 20 M gas limit on the Raspberry Pi 4.

5.5.3. Energy consumption

In order to perform an energy consumption analysis of PoW and PoA for the selected resource-constrained devices, highly accurate hardware was used to carry out measurements on the two SBC models evaluated. Specifically, the energy consumption measurements were performed with a Joulescope [69], which is a high-precision power meter with a 1.5 nA resolution that is able to measure voltage and current at 2 million samples per second with a 250 KHz bandwidth.

During the energy consumption tests, different amounts of workload were generated for both SBCs. In the case of the Orange Pi, since it is less powerful, only one peer was included for both PoA and PoW. For the Raspberry Pi 4, two tests were performed for comparing PoA and PoW energy consumption: one for two peers and another one with eight peers.

Figures 23, 24 and 25 show the energy consumption obtained by the SBCs for PoW and PoA. In particular, Figure 23 shows an example of the evolution of the energy consumption on the Orange Pi during the execution of the PoA algorithm. In such a Figure different noticeable peaks with similar maximum values and duration can be observed, which correspond to the performed block validations. For example, the interval defined as P1 in Figure 23 (delimited by two vertical red dashed lines), contains a peak that has a duration of approximately 1.8 s and an average current consumption of 600 mA.

Figure 24 shows a comparison between PoW and PoA when executing two peers on the Raspberry Pi 4. In this case, it can be first observed the considerable difference in consumption between PoW and PoA. It can be also observed that the performance exhibited by the Raspberry Pi significantly surpasses the one delivered by the Orange Pi: the Raspberry Pi is able to validate blocks approximately every 4 s, compared to the 10 s needed by the Orange Pi. Moreover, the time required by every validation is considerably smaller (about 250 ms, as indicated through interval P1), which is due to the computational power difference between both SBCs.

Regarding the energy consumption difference between PoW and PoA, Figure 24 allows for observing that PoW is significantly more energy-intensive. In fact, it can be stated that PoW is not suitable for low computational power devices like the Orange Pi. It is also interesting to observe through Figure 24 that there is an idle interval of approximately one second (labeled as P2), which corresponds to the search of PoW for new peers prior to initiating the mining operation. The rest of the time PoW energy consumption remains constant, with values ranging between 0.9 and 1 A.

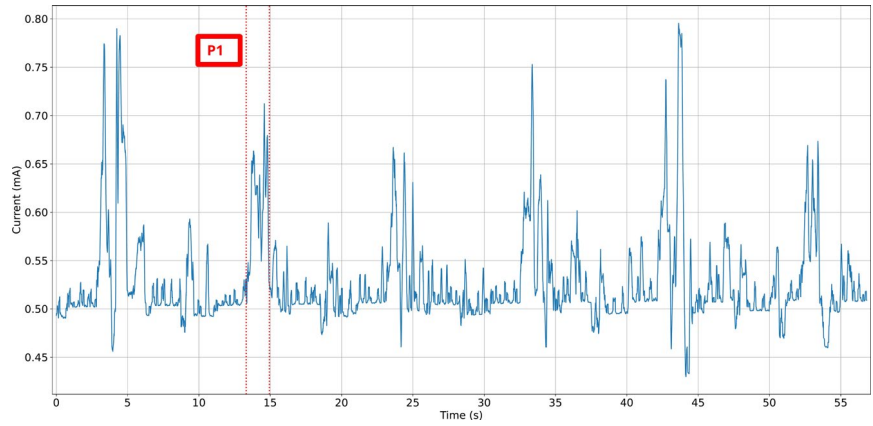

**Figure 23.** Power consumption for the PoA algorithm on the Orange Pi by simulating one peer.

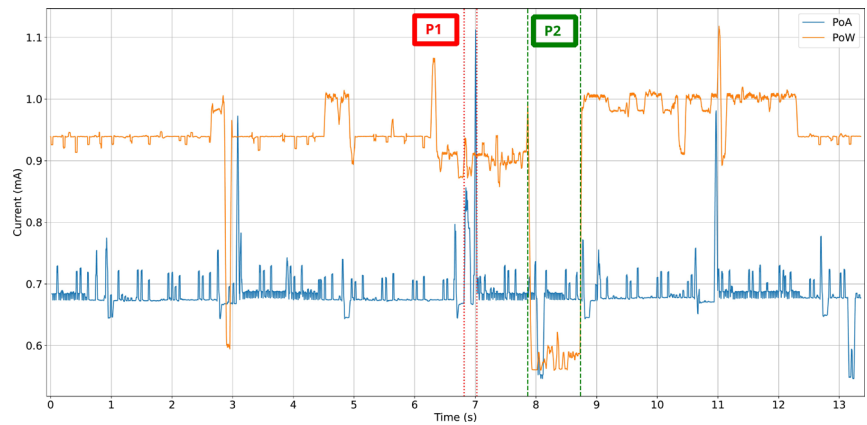

**Figure 24.** Power consumption for the PoA vs PoW algorithm on the Raspberry Pi by simulating two peers.

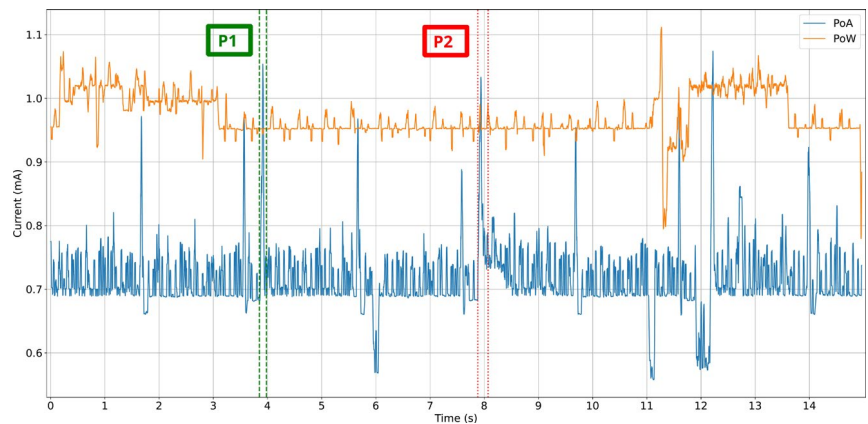

**Figure 25.** Power consumption for the PoA vs the PoW algorithm on the Raspberry Pi by simulating eight peers.

The increase in the number of peers exhibits a comparable trend in both algorithms. The main difference is a slight increase in their average consumption. As illustrated in Figure 24, for PoA with two peers, the average consumption was 0.68 mA, while only 0.72 mA were consumed on average for eight peers (this is shown in Figure 25). In the case of PoW, 0.92 mA were consumed for two peers and 0.96 mA for eight peers.

Finally, regarding PoA performance, it can be observed in Figure 25 that, despite increasing from 2 to 8 peers, the obtained time intervals are similar (approximately 220 ms, as illustrated by P1 and P2 in such a Figure), with validations occurring every 4 seconds.

## 6. Key Findings on the Experiments

After the theoretical and practical analyses performed for this article, several key findings can be highlighted:

- The read and write operation latency of the proposed decentralized system can be really low, in the same order as similar centralized applications but providing higher levels of cybersecurity. This fact indicates that applications for scenarios where low response times are not critical can benefit from the use of blockchain, since the difference in terms of user experience is negligible.
- On the same conditions, the use of PoA significantly increases TPS with respect to the use of PoW. Therefore, the use of PoA includes benefits in terms of efficiency, energy consumption and speed.
- Setting an appropriate gas limit is essential when it comes to determine the blockchain behavior. As it was observed in the previously described experiments, increasing the gas limit significantly increases TPS, since blocks can include more transactions. However, this requires nodes to have more computational resources to verify these transactions, and it would take longer to propagate blocks through the network (i.e., more overhead would be involved). Therefore, resource-constrained devices would not be able to validate blocks on time, losing the capacity to keep the network synchronized and eventually leading to centralization.

  According to the results of the experiments presented in this article, both for low and high gas limits, and even with high transaction loads (e.g., 250 transactions per second is considered a high load for the application described in this paper), the system maintained acceptable request rates. Therefore, low gas limits are effective in terms of performance and increase decentralization, which is essential when using consensus protocols such as PoA.
- With respect to CPU usage and CPU power consumption, the behaviour of PoW substantiates the one described in [59], where the plateau of PoW is associated with the constant mining process for securing the blockchain. In any case, it must be noted that the previously described CPU usage and power consumption results are actually unlikely in practice, since the simulated workloads (i.e., a progressive load that ranged progressively from 100 to 1,000 transactions) were aimed at testing the performance of the system rather than to recreate a realistic scenario for the application described in this article.
- As for the deployment in SBC devices, its use can be beneficial for the greater proximity to the end users. The experiments performed confirm that the use of SBCs is viable for implementing both PoA and PoW-based blockchain nodes, but it is necessary to consider the limitations of such devices in terms of computational power to make proper use of the system and to not to degrade their performance. Specifically, the experiments showed that, to provide an adequate performance level, devices required at least 1 GB of RAM (combined with at least 1 GB of swap memory) together with a CPU with at least 4 cores of a relatively powerful ARM processor.

## 7. Conclusions

This article presented the design, implementation and evaluation of a solution for recording and verifying academic records through a DApp deployed on a Ethereum blockchain. The application uses a decentralized storage system based on IPFS to manage off-chain data. The proposed implementation is open-source (under GPL-3.0 license), so it can serve as a reference for future researchers and developers to carry out further improvements and tests.

To demonstrate the efficiency of the developed solution, it was evaluated in terms of performance (transaction latency and throughput) and efficiency (CPU usage and energy consumption) on both traditional computers and SBCs. For traditional computer-based nodes and normal workloads (under 250 requests per second), read-operation latency was approximately 1 ms, while such a latency increased to roughly 5 ms for a Raspberry Pi 4, but to almost 10 s in some situations for the least powerful SBC evaluated (an Orange Pi One). When comparing the results between PoA and PoW, it is clear that the PoA protocol is more environmentally friendly and requires less CPU load for both traditional computers and SBCs. Moreover, the impact of Ethereum's gas limit was analyzed, observing its clear influence on the system performance.

**Author Contributions:** Conceptualization, T.M.F.-C.; methodology, P.F.-L. and T.M.F.-C.; investigation, G.F.-B., I.F.-M., P.F.-L. and T.M.F.-C.; writing—original draft preparation, G.F.-B., I.F.-M. and T.M.F.-C.; writing—review and editing, G.F.-B., I.F.-M., P.F.-L. and T.M.F.-C.; supervision, P.F.-L. and T.M.F.-C.; project administration, T.M.F.-C.; funding acquisition, T.M.F.-C. All authors read and agreed to the published version of the manuscript.

**Funding:** This work has been funded by grant TED2021-129433A-C22 (HELENE) funded by MCIN/AEI/10.13039/501100011033 and the European Union NextGenerationEU/PRTR.

**Institutional Review Board Statement:** Not applicable

**Informed Consent Statement:** Not applicable

**Conflicts of Interest:** The authors declare no conflicts of interest.